\documentclass[10pt]{iopart}
\usepackage{iopams}
\usepackage{graphicx}
\usepackage{multicol}
\usepackage{color}
\usepackage{cite}
\def\figref#1{Figure~\ref{#1}}
\def\secref#1{Section~\ref{#1}}
\def\tabref#1{Table~\ref{#1}}

\begin{document}

\title[Microwave emission by nonlinear crystals]{Microwave emission by  nonlinear
crystals irradiated with a high-intensity, mode-locked laser}

\author{A F Borghesani$^{1}$, C Braggio$^{2}$ and M Guarise$^{2}$}
\address{$^{1}$CNISM Unit,  Department of Physics and Astronomy, University
of Padua
\\
 and Istituto Nazionale di Fisica Nucleare, Sez. Padova \\ Via F. Marzolo
8, I-35131 Padua, Italy}
\address{$^{2}$Department of Physics and Astronomy, University of Padua \\
and Istituto Nazionale di Fisica Nucleare, Sez. Padova \\ Via F. Marzolo
8, I-35131 Padua, Italy}
\ead{armandofrancesco.borghesani@unipd.it}
 
\begin{abstract} 
We report on the experimental investigation of the efficiency of
some nonlinear crystals to generate microwave (RF) radiation  as a result of optical rectification (OR) when irradiated with intense
pulse trains delivered by a mode-locked laser at $1064\,$nm. We have investigated lithium triborate (LBO), lithium niobate (LiNbO$_3$), zinc selenide (ZnSe),
and also potassium titanyl orthophosphate (KTP) for comparison with previous
measurements.
The results are in good agreement with
the theoretical predictions based on the form of the second-order nonlinear
susceptibility tensor. For some  crystals we investigated also the second
harmonic generation (SHG) to  cross check the theoretical model. We  confirm
the theoretical prediction that OR 
leads to the production of higher order RF harmonics that are overtones
of the laser repetition rate.
\end{abstract}


\vspace{2pc}

\vspace{2pc}
\submitto{\JOPT}
\maketitle

\ioptwocol

\section{Introduction}\label{sect:intro} 

Nonlinear Optics is a well established branch of Physics~\cite{yariv,bloembergen}.
The discovery of materials with nonlinear optical properties has paved the
way to the
vast realm of optical frequency conversion. Among the most important applications
there is second harmonic generation (SHG). Also optical rectification (OR)
in nonlinear electro-optic crystals is a  well
known phenomenon~\cite{franken1961,bass1962,bass1965} that has been exploited
for the production of (sub)picosecond, microwave and terahertz bandwidth
radiation as a consequence of
frequency mixing~\cite{yang1971,xu1992,zhang1992b,rice1994}. 

Microwave (RF) pulses are produced by several techniques based on optical
heterodyning~\cite{bridges1972,yao2009,khan2010}. The beating between two
nearby laser lines or between the various Fourier components of the optical
spectrum of an ultrashort laser pulse produces optically rectified electrical
pulses at the difference frequency in the microwave- and far infrared range
(from GHz up to THz)~\cite{niebuhr1963,lengfellner1987,Nahata1996,nahata1998}.

In a recent Letter~\cite{Borghesani2013}, we have reported on a different
way to produce long microwave pulses by irradiating a second-order nonlinear
KTiOPO$_4$ (KTP) crystal with $500\,$ns-long  pulse trains delivered by a
high-intensity, mode-locked laser in the near infrared at $1064\,$nm. The
$10\,$ps-long pulses are repeated at a rate $f_0\approx 4.6\,$GHz.   
 As a result, the laser spectrum is quite pure and optical rectification
is obtained by only exploiting the high strength of the laser electric field.

The fast electronic, second-order response of the nonlinear crystal gives
origin to a time-dependent polarization $P(t)$ that closely follows the envelope
$ \vert E_0 (t)\vert^2$ of the intense laser pulses electric field $E(t)=E_0(t)\cos{\omega_L
t},$ with $\omega_L =2\pi f_L$ where $f_L$ is the laser frequency~\cite{graf2000}.
The low-frequency branch of  the Fourier spectrum of $P(t)$ contains the
fundamental RF\ harmonic at  frequency $f_0$ and several of its overtones.
The fundamental harmonic can be easily detected by placing the crystal in
a microwave cavity that acts as a narrow bandpass filter. Its amplitude dependence on the crystal orientation with respect to the laser beam polarization
gives pieces of information on the crystallographic structure of the sample.
This kind of technique can also be used as an inline tool to monitor the
stability of a high repetition rate, mode-locked laser. The possible advantages
of such a technique have recently been highlighted~\cite{Braggio2014}. 

The technique reported in the  previous Letter could be also exploited to
measure the elements of the nonlinear second-order optical tensor of a given
crystal in addition to the Maker-fringe technique\cite{maker1962} and it
additionally gives the researchers the opportunity to produce microwave overtones
in a controlled way. 

In order to test these statements, we have carried out additional measurements
of RF generation in several crystals,{ which were never previously investigated to this purpose}, whose crystallographic properties differ
from those of KTP,
and in KTP itself for comparison sake.  We have chosen
lithium triborate (LiB$_3$O$_5$ or, simply,  LBO), lithium niobate (LiNbO$_3$),
and zinc selenide (ZnSe) because the elements of their second-order nonlinear
optic tensor are relatively well known. The measurements have been carried out by placing the crystals either in
a microwave cavity or in a waveguide of much wider passband in order to measure
the microwave overtones. As RF generation  is a second-order nonlinear phenomenon
intimately related to optical SHG, optical second harmonic (SH) light has
simultaneously been measured as a cross check of the validity of our approach.

The paper is organized as follows. In Sect.~\ref{sect:expapp} the details
of the experimental setup are given. In Sect.~\ref{sect:th} the theoretical
basis for the understanding of the phenomenon is described. The results are
presented and discussed in Sect.~\ref{sect:res}. Finally, the conclusions
are drawn in Sect.~\ref{sect:conc}.

\section{Experimental Setup}\label{sect:expapp}
\begin{figure}[htbp]
    \centering
\includegraphics[angle=0,width=\columnwidth]{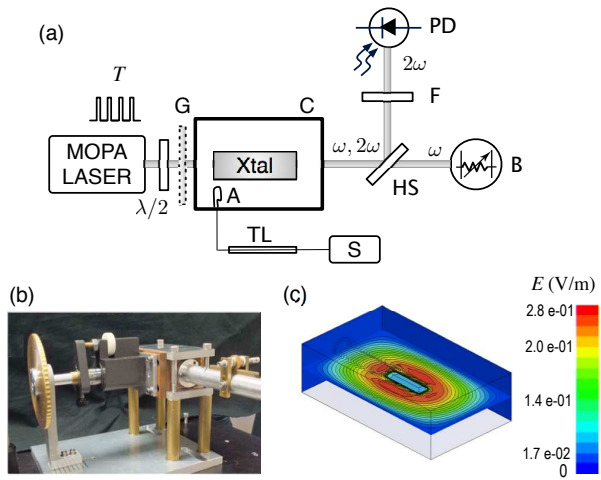}
    \caption{\small { Scheme of the experimental apparatus for RF cavity
measurements (color online). (a) $\lambda/2$: rotary half-wavelength plate,
G: rotary goniometer for the crystal rotation, C: RF cavity, XTAL: crystal,
A: RF pick-up antenna, TL: transmission line, S: oscilloscope, HS: harmonics
separators, F: infrared filter, B: bolometer. (b) Photo of the cavity assembly:
on the left, we see the  rotary crystal holder wheel. The laser beam enters
the cavity through the hollow shaft connected to the wheel. (c) Electric
field distribution around the crystal in the cavity.}\label{fig:scheme}}
\end{figure}

{ We show in~\figref{fig:scheme} the scheme of the apparatus for the measurements in the RF cavity. The apparatus for the measurements in the waveguide is much simpler and does not deserve description.}
 The experimental setup and technique have thoroughly been described elsewhere.
We briefly recall here  the main characteristics of the apparatus while referring
to literature for details~\cite{Borghesani2013,borghesani2014,Braggio2014}.  

A home-made, infrared ($\lambda=1046\,$nm), mode-locked laser of high intensity
delivers a $500\,$ns-long train { ($T$)} of $N\approx 2300,$ $10\,$ps-long pulses~\cite{agnesi2011} { (see \figref{fig:scheme}, part (a))}.
The repetition rate $f_0\approx 4.6\,$GHz used in our experiment is not a limiting
factor as the fast electronic response of the crystals makes them responsive
to lasers of much higher repetition rate~\cite{krainer2002,keller2003}. The
laser beam is linearly polarized along the $\hat y$ direction and propagates
along the $\hat z $ direction. A half-wave plate { ($\lambda/2$)}, mounted on a rotary goniometer in the beam path, can be rotated through an angle $\theta$ to rotate the beam polarization through an angle $2\theta$ with respect to the original direction.
Then, the beam impinges onto the entrance face of the crystal. The laser
beam has an ellipsoidal Gaussian profile with effective area $S\approx 2.6\,$mm$^2$ { and its intensity averaged over the spot size is $I.$} It can be reduced by inserting calibrated neutral density filters
in the beam path and is measured with a bolometer { (B)} (Coherent, mod. J-25-MB-IR). { A picture of the cavity assembly is shown in ~\figref{fig:scheme}, part (b).}

The crystals under investigation, of typical size $4\times 4\times 10\,$mm$^3,$
are cut in the shape of a right square prism with the long edge directed
along the $\hat z$ direction and the square face lying in the  $(\hat
x, \hat y)$ plane. 
In order to investigate the fundamental RF harmonic, the crystal is placed
in a rectangular RF cavity designed so as to sustain a TE$_{101}$ mode tuned
to the laser repetition rate.{ A contour plot of the electric field distribution around the crystal in the cavity is shown in~\figref{fig:scheme}, part (c).} If higher RF harmonics are to be detected,
the crystal is placed in a coaxial waveguide consisting of two concentric
cylindrical conductors of inner and outer diameter $3.9\,$mm and $9.1\,$mm,
respectively, that is designed to support TEM modes up to $\approx 12\,$GHz
and TE and TM modes of higher frequency~\cite{pozar,borghesani2014}. 

\begin{figure}[htbp]
    \centering
\includegraphics[angle=180,width= \columnwidth]{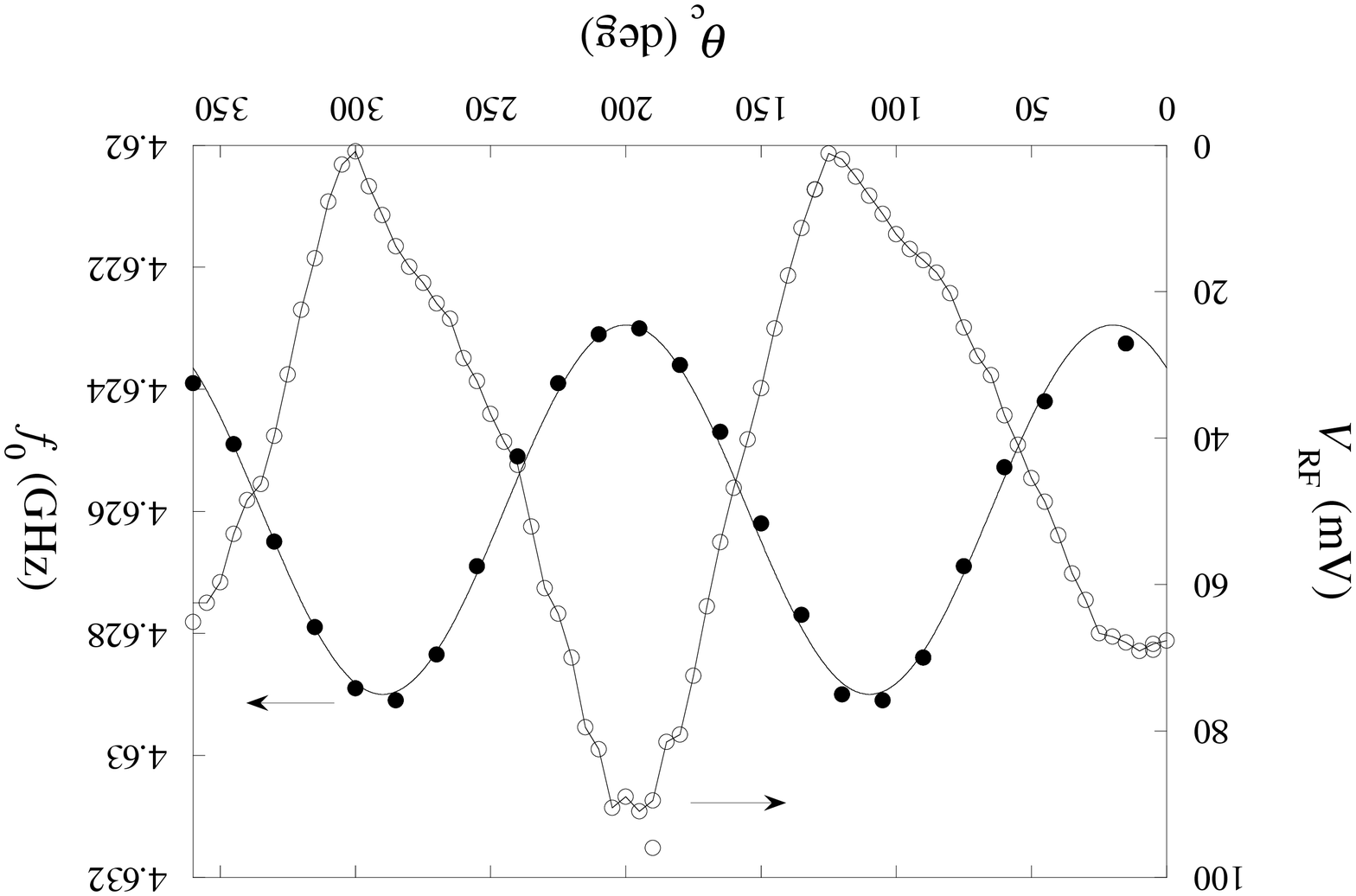}
    \caption{\small RF amplitude $V_\mathrm{RF}$ (open symbols, left scale)
and cavity resonance frequency $f_0$ (closed symbols, right scale)  vs $\theta_c$
for KTP. The lines are  only a  guide for the eyes.\label{fig:KTPf0Vrf}}
\end{figure}

When the cavity is used, the crystal is mounted on a rotary goniometer { (G)} and
can be rotated through an angle $\theta_c$ in order to maximize the RF signal
detected by a critically coupled antenna.
The electrical power transferred to the cavity field is $\propto \vert \mathbf{E}_\mathrm{RF}\cdot
\mathbf{j}\vert $, where $\mathbf{j}$ is the polarization current density\cite{jackson}.
Thus, we expect that RF signal shows two maxima and two minima for a complete
turn of the shaft about its $\hat z$ axis as is shown in \figref{fig:KTPf0Vrf},   
in which we plot how the RF signal amplitude $V_\mathrm{RF}$
of the KTP crystal and the resonance frequency $f_0$ of the cavity vary with
the crystal alignment with respect to the cavity mode polarization. Similar
results are obtained with all other crystals.
 We note that a crystal rotation
about one of its geometrical symmetry axis can only maximally align the polarization
along the cavity mode polarization because the geometrical axes may not coincide
with the crystallographic ones. { We believe that this misalignment is responsible for the asymmetry of $V_\mathrm{RF}$ as a function of the angle $\theta_c$.}
We further note that $f_0$ is the lowest
when $V_\mathrm{RF}$ is the largest. 

 The light exiting the opposing crystal face contains both contributions
of the pump laser and of the second harmonic (SH) and exits the cavity through
a small opening. The infrared component is filtered out by means of a suitable
combination of harmonics separators { (HS)} and bandpass filters { (F)} so that the SH component
can be measured by a photodiode { (PD)}.
When the waveguide is used, no provisions are made to optimize the crystal
orientation and to measure the SH. 

The RF signal { is picked up  by a small antenna (A)} and its amplitude is so large that it can directly be observed with
the oscilloscope { (S)} (LeCroy, mod. WaveRunner 6000A for the cavity- and LeCroy,
mod. LabMaster/SDA/DDA 8Zi-B for the waveguide measurements), except for
the LBO crystal, for which we used a $38\,$dB-gain microwave amplifier (Miteq,
mod. AMF-4D-001120220-10P).

The amplitude of the RF emission is obtained by fitting a Lorentzian function
to the numeric Fourier transform of the recorded RF signal. The conversion
factor between the Lorentzian amplitude  and the actual rms RF amplitude
is simply $1/\sqrt{M/4\pi},$ where $M$ is the number of points sampled by
the oscilloscope. 

\section{Theoretical background}\label{sect:th}

The observed phenomenon is explained by introducing the  crystal second-order
nonlinear optical susceptibility tensor of elements $d_{ijk}.$ 
The laser electric field after the half-wave plate at the crystal entrance
face can be written as
\begin{equation}
\mathbf{E}(t) = E_0 \mathbf{\hat e}\cos{\left(
\omega_L t\right)} u(t, \Delta, \tau,N)
\label{eq:e}\end{equation}
in which $E_0$ is its amplitude and $ \mathbf{\hat e}$ is its polarization
vector given by
\begin{equation} 
\mathbf{\hat e}=\left(\begin{array}{c}\cos{2\theta}  \\ \sin{2\theta}   \\0\end{array}\right)
\label{eq:epol}\end{equation}
The function $u$ represents the pulse train and can be written as
\begin{equation}
u(t,\Delta,\tau,N)=
 \sum\limits_{m=0}^{N-1}\left[
H(t-m\Delta)- H(t-m\Delta-\tau)
\right]
\label{eq:train}\end{equation}
in which the pulses are assumed rectangular.
 $H$ is the Heaviside step function, $\Delta=f_0^{-1}$ is the time interval
between pulses, $\tau\approx 10\,$ps is the pulse duration, and $N$ is the
number of pulses in the train. $f_0$ is the pulses repetition rate and, thus,
the fundamental RF harmonic. Actually, the pulses have a $\mathrm{sech}^2$
 shape but this fact does not affect the conclusions we draw.

Limiting our attention to second-order phenomena, the cartesian components
of the nonlinear polarization produced by the strong laser field are
\begin{equation}
P_i  = 2\sum\limits_{jk}d_{ijk}E_jE_k 
\label{eq:pol}\end{equation}By introducing the time dependence of the field,
we get for the polarization
\begin{eqnarray}
P_i(t)&=& E_0^2 \left[
1+\cos{\left(2\omega_L t\right)}\right]u(t,\Delta,\tau,N)\sum\limits_{jk}
d_{ijk}\hat e_j\hat e_k\nonumber\\
&=& P_i^\mathrm{OR}+P_i^\mathrm{SHG}
\label{eq:pt}\end{eqnarray}
The first term in square brackets gives the OR contribution whereas the second
one is responsible for the SHG. The structure of the quasistatic polarization due to OR  is
\begin{equation}
P_i^\mathrm{OR}(t)=E_0^2u(t,\Delta,\tau,N)\sum\limits_{jk} d_{ijk}\hat e_j
\hat e_k
\label{eq:por}
\end{equation}
The factor $E_0^2$ is proportional to the laser intensity $I.$ Owing to the
presence of the function $u(t,\Delta,\tau,N),$ the Fourier spectrum of $P_i^\mathrm{OR}$
is a comb of regularly spaced frequencies that are integer multiples of the
fundamental harmonic $f_0$~\cite{cundiff2003}. In our experiment, $4.6\,\mathrm{GHz}\le
f_0 \le 4.7\,$GHz and the spectrum extends roughly up to  a frequency $f_M=
\tau^{-1}\approx 100\,$GHz. 
Finally, the sum involving the elements of the second-order nonlinear optical
susceptibility tensor and the components of the vector $\mathbf{\hat e}$
must in general be a function $h_i$ of the rotation angle $\theta$ of the
half-wave plate of the form
\begin{equation}
h_{i}\equiv h_i(4\theta, d_{ijk})= \sum\limits_{jk} d_{ijk}\hat e_j \hat
e_k
\label{eq:hi}\end{equation}
The 4-fold periodicity stems from the quadratic terms $\hat e_j \hat e_k.$
The actual analytic form of $h_i$ depends on the crystals via the explicit
form of $d_{ijk}$ but it is time independent
 and does not modify the spectral composition of $P_i^\mathrm{OR}$ that can
be thus expanded in Fourier series 
\begin{equation}
P_i^\mathrm{OR}(t)= I h_i\left(4\theta, d_{ijk}\right)\sum\limits_l p_l \cos{(2\pi
lf_0 t)}
\label{eq:pfour}\end{equation}
Here, the amplitude coefficient $p_l$ of the $l$-th RF harmonic depends on
the Fourier transform of the actual shape of $u$.

It can be shown~\cite{jackson,griffiths} that, in the condition of our experiment,
the amplitude of the RF field radiated by the crystal is proportional to
the second time derivative of the slowly varying dielectric polarization

\begin{equation}
{\ddot{P}}_i^\mathrm{OR}
 \propto I h_i\left(4\theta, d_{ijk}\right) f_0^2\sum\limits_l l^2p_l \cos{(2l\pi
f_0t)}
\label{eq:d2p}\end{equation}
Thus, the RF signal detected in our apparatus is proportional to it. All
harmonics should be  proportional to the laser intensity and  share the same
angular behaviour.

Similarly, the spectrum of the SH radiation emitted by the crystal, which
is the time-averaged square modulus of the second time derivative of $P_i^\mathrm{SHG},$
consists in a restricted-band frequency comb centered around the frequency
$2 f_L\approx 5.64\times 10^{14}\,$Hz and width of order $\tau^{-1}\approx
100\,\,$GHz$\,\, \ll 2f_L$ so that all frequencies are in the optical region.
As a result, the integrated intensity $I_\mathrm{SH}$ can be cast in the
form
\begin{equation}
I_\mathrm{SH} \propto \langle\left\vert
\left(\ddot{P}_i^\mathrm{SHG}\right)\right\vert^2\rangle\propto 
I^2 v_i\left(8\theta,4\theta,d_{ijk}\right)
\label{eq:ish}\end{equation}
in which 
\begin{equation}
v_i \equiv v_i \left(8\theta,4\theta,d_{ijk}\right) =\sum_{jklm} d_{ijk}{d_{ilm}}\hat
e_j \hat e_k \hat e_l \hat e_m
\label{eq:vi}\end{equation}
 The 8- and 4-fold periodicity stems from the quartic terms in $\hat e$.
 
We note that the elements of the nonlinear second-order susceptibility tensor
for OR, $d^0_{ijk},$ and for SHG, $d^{2\omega_L}_{ijk},$ could in principle
be different but we have shown that in this experiment they actually are
equal so that no superscript is needed to tag  them~\cite{Borghesani2013}.

Finally, as a consequence of the symmetry with respect to the interchange
of indices $j$ and $k$ in ~\eref{eq:pol}, we will use  the  contracted indices
form of the optical tensor that is  represented by a $3\times 6$ matrix of
elements $d_{il}$ that operates on the $E^2$ column vector to yield the amplitude
of the second-order nonlinear polarization~\cite{yariv}.

\section{Experimental Results and Discussion}\label{sect:res} 
In this section we present and discuss the results obtained about the efficiency
and properties of the microwave produced by several nonlinear crystals in
both  microwave cavity and waveguide. 
We anticipate that the results obtained with the two different experimental
setups are consistent with each other.

\subsection{Materials}\label{sect:mat}
We have investigated the following crystals: LBO, LiNbO$_3,$ ZnSe, and  KTP. 
{ As this experiment is not aimed at SHG, the crystals are not specifically cut to obtain phase matching.} Their crystallographic properties are summarized in \tabref{tab:xtals}. 
\begin{table}[!h] 
\caption{\label{tab:xtals}\small Crystallographic properties}

\begin{indented}
\item[]\begin{tabular}{@{}llll} \br
Crystal& Structure & Group & Nonzero $d_{{il}}$ elements \\ \mr
LBO & orthorombic &$ 2mm$ &  $d_{15},d_{24},d_{31},d_{32}
, d_{33}$ \\ 
LiNbO$_{3}$& trigonal & $R3c$ &  $d_{15},d_{16},d_{21},d_{22}, d_{24}$\\
&   &  & $d_{31},d_{32} ,d_{33}$\\
ZnSe & cubic zincblende & $\bar 4 3 m$ & $d_{14},d_{25},d_{36}$  \\
KTP& orthorombic & $2mm$ & $d_{15},d_{24},d_{31},d_{32}
, d_{33}$ \\ \br
\end{tabular}
\end{indented}
\end{table}

\noindent For KTP, only three out of the five nonzero coefficients are independent
as $ d_{15}=d_{31}$ and $d_{24}=d_{32}.$ 
Their values are $d_{15}=1.73$,  $d_{24}=3.45$, and $d_{33}=13.5\,$~\cite{pack2004}.
The tensor elements of LBO are smaller than those of KTP and their values are $d_{15}=-0.897$, $d_{24}=0.958$, $d_{31}=-0.854$, $d_{32}=0.992$, and
$d_{33}=0.057$~\cite{lin1990}. For
LiNbO$_3,$ only three tensor elements are independent as $d_{15}=-d_{24}=d_{31}=d_{32},$
$d_{16}=d_{21}=-d_{22},$ and $d_{32}= d_{31}.$ Their values are  $d_{15}=-4.3,$ $d_{16}=-2.1,$ and~$d_{33}=-27$~\cite{roberts1992}. All values are expressed in pm/V.  Finally, ZnSe, though optically isotropic, is not endowed with
an inversion center and its only three non vanishing tensor elements are
equal to each other with value $d_{14}=d_{25}=d_{36}=33\,$pm/V~\cite{kuhnelt1998,wagner1998}. We have to note that the spread of the values reported in literature is quite
large.

\subsection{RF generation efficiency}\label{sect:RFeff}
According to~\eref{eq:d2p}, the amplitude of the signal detected by the antenna
in the cavity or measured in the waveguide, $V_\mathrm{RF},$ is proportional
to the laser intensity $I$ for a fixed direction $\theta$ of the laser beam
polarization and at constant laser repetition rate $f_0.$

\begin{figure}[htbp]
\centering
\includegraphics[width= \columnwidth,angle=180]{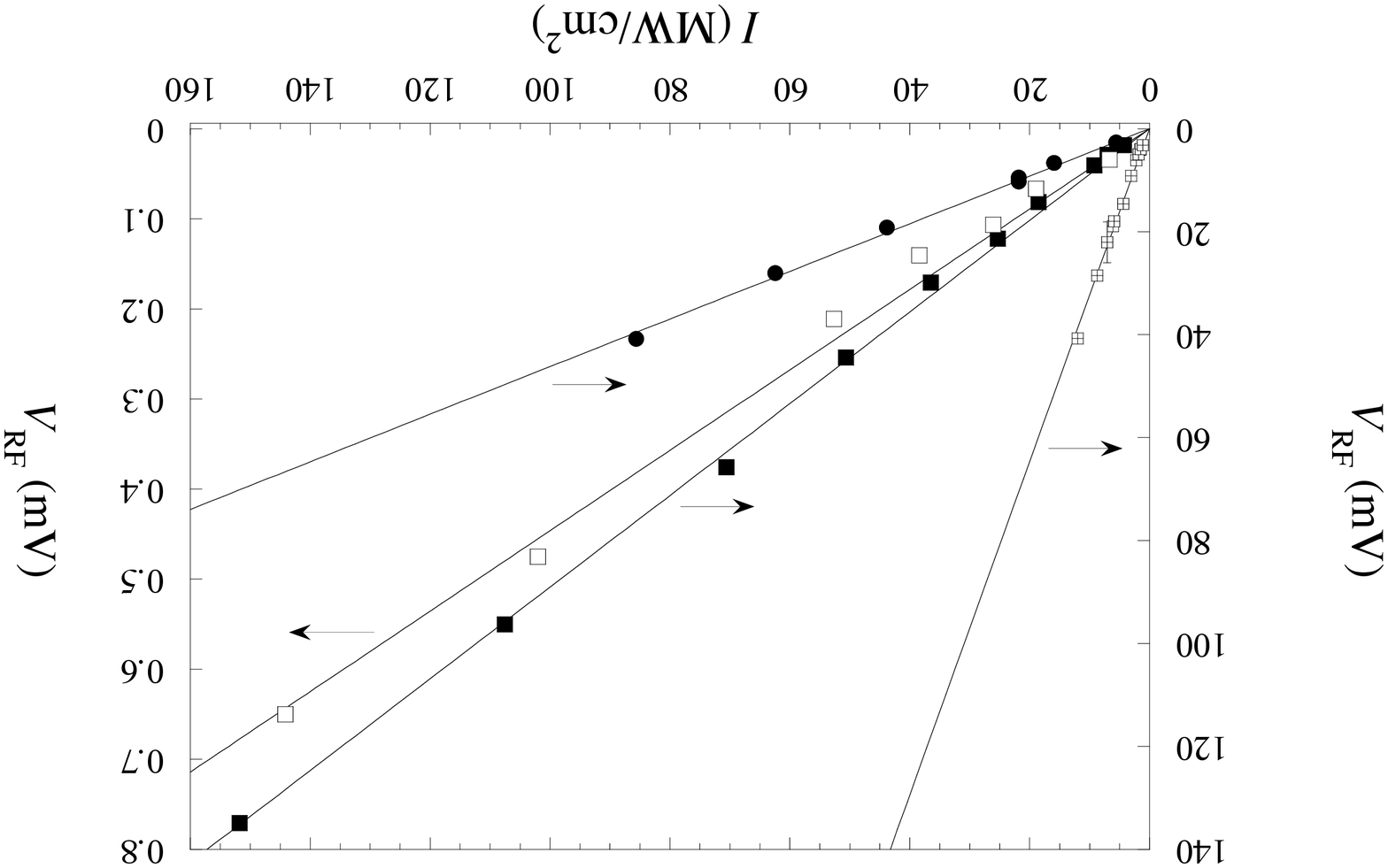}
\caption{\small 1st RF harmonic amplitude $V_{\mathrm{RF}}$ vs $I.$ Left
scale: ZnSe (crossed squares), LiNbO$_{3}$ (closed circles), and KTP (closed
squares). Right scale: LBO (open squares). The dot size is comparable to
the error bars.
\label{fig:VRFvsI_4XTALS}}
\end{figure}

We report  in~\figref{fig:VRFvsI_4XTALS} the linear relationship between
$V_\mathrm{RF}$ and $I$ measured for all crystals. 

\noindent
For ZnSe we used a weaker \(I\) because of its smaller damage threshold.
The observed linearity confirms that the generation of RF is a second-order
nonlinear effect.  We note that ZnSe is the
most efficient crystal and LBO is the least efficient one.
ZnSe is $\approx 3.6 $ times more efficient than KTP and $ \approx 7$ times
more efficient than LiNbO$_3,$ and, finally, $\approx 700$ times more efficient
than LBO.
Qualitatively, this result mirrors the value of the largest tensor element
of each crystals. The largest coefficient is that of ZnSe, the smallest belonging
to LBO. Unfortunately, the slope of the $V_\mathrm{RF}-I$ relationship is
not related in a simple way to the elements $d_{il}$ of the optical tensor
so that no easy quantitative comparison is possible with a theoretical prediction~\cite{Borghesani2013}.

\begin{figure}[htbp]
\centering
\includegraphics[width=1.4 \columnwidth,angle=90]{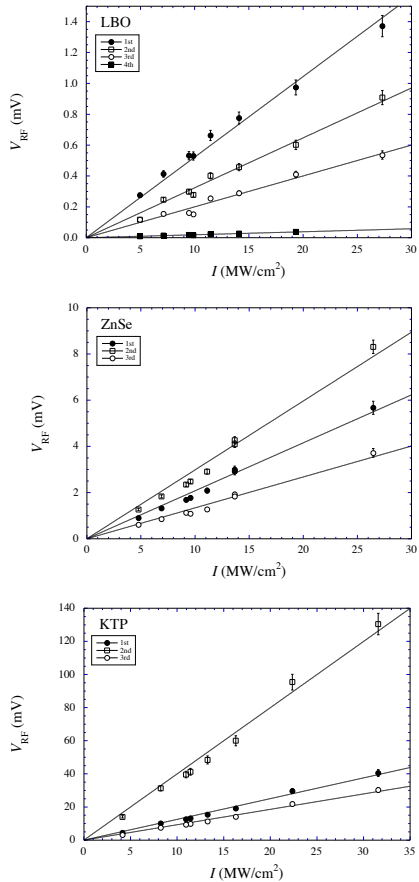}
\caption{\small $V_{\mathrm{RF}}$ vs $I$ for LBO, ZnSe, and KTP (from top)
for the harmonics
observed in the waveguide. The 4th harmonics in KTP is too small to be shown.\label{fig:VRFvsIL3XtalAllHarm}}
\end{figure}

 In the waveguide, we have been able to detect several higher order RF
harmonics, up to four for KTP and LBO and up to three for ZnSe, as shown
on~\figref{fig:VRFvsIL3XtalAllHarm}.
 LiNbO$_3$ was not investigated in this
experiment. The amplitude of all harmonics depends linearly on $I,$ thereby
validating~\eref{eq:d2p}. We note that the harmonics are not naturally ordered according to their expected
strength. In KTP and ZnSe the second harmonic is stronger than the first
one. Moreover, KTP appears to produce more RF than ZnSe in contrast with
the cavity result. In any case, LBO is confirmed to be the least efficient.
We noted that the placement of the crystal in the waveguide is a critical
issue. Tiny displacements of the same crystal could lead to significant relative
and absolute changes of the harmonic strengths. The reason for this behaviour
might be ascribed to the strong non uniformity of the electric field distribution
in the waveguide~\cite{pozar} that makes the positioning of the crystal hardly
reproducible.

{  As a final remark, we note that, in the present experiment, the RF generation is not affected by the SHG because the efficiency of SHG is well below $1\,\%$~\cite{Braggio2014} owing to the fact that the crystals are not cut so as to obtain phase matching at the wavelength of the experiment. Thus, the pump beam is not expected to significantly be depleted by SH conversion.}

\subsection{SHG efficiency}\label{sect:SHGeff}
\begin{figure}[htbp]
\centering
\includegraphics[width= \columnwidth,angle=180]{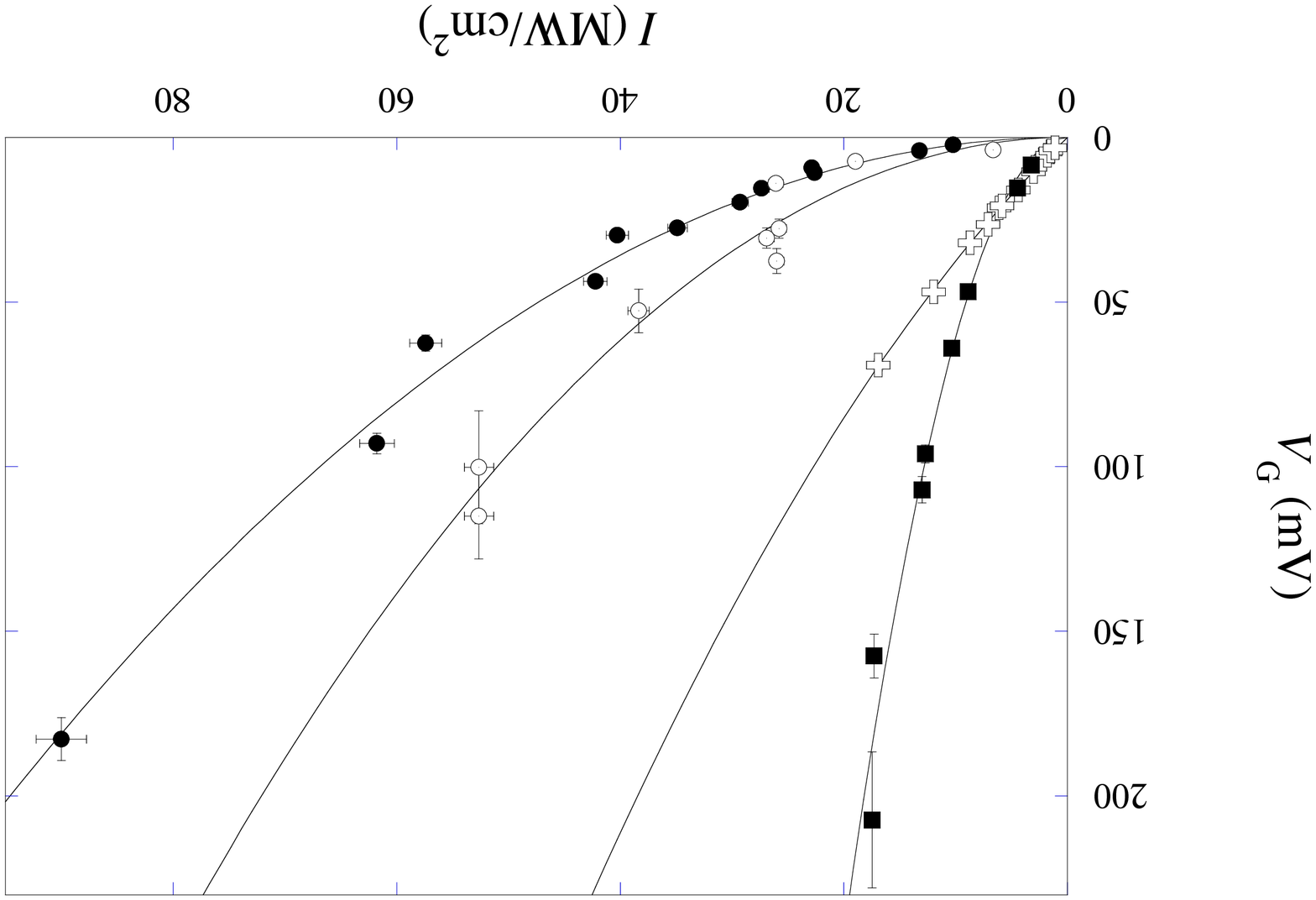}\caption{\small
SH intensity $V_\mathrm{G}$ vs $I:$   LiNbO$_{3}$  (squares), ZnSe (crosses),
LBO\ (open circles), and KTP (closed circles). \label{fig:VgvsIlallXTALSbis}}\end{figure}

As previously mentioned, we have also carried out SHG measurements when using
the microwave cavity. The SH intensity is measured with a photodiode whose
output is averaged over the duration of
the laser pulse train, thus yielding $V_\mathrm{G} $. All investigated crystals
do actually emit light at $532\,$nm with different efficiency, as shown in~\figref{fig:VgvsIlallXTALSbis}.

\noindent As expected, according to \eref{eq:ish}, the SH intensity is a quadratic
function of the laser intensity $I. $ KTP is the less efficient of all of
them whereas LiNbO$_3$ is the most efficient one.

\subsection{RF dependence on the laser polarization}\label{sect:RFang}
As observed in our previous experiment\cite{Borghesani2013}, and explained
in \secref{sect:th}, the microwave emission depends on the crystal axes orientation
with respect to the laser polarization. This fact can be exploited to build
an experimental setup for the measurement of the elements of the second-order
optical tensor. To this goal, the measurements can be either carried out
in a cavity or in a waveguide.  
\begin{figure}[htbp]
    \centering
\includegraphics[angle=180,width= \columnwidth]{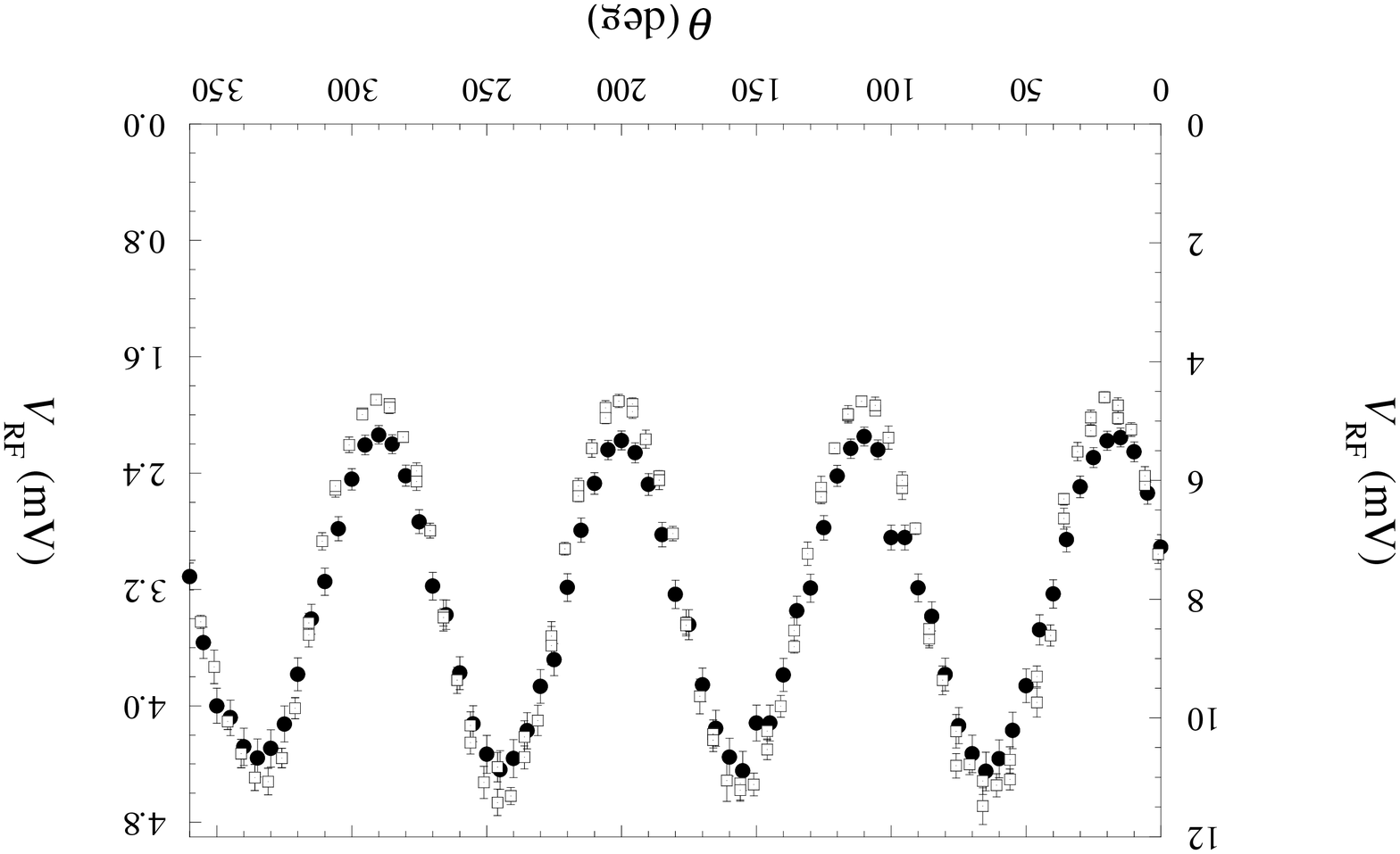}
    \caption{\small Comparison of the angular dependence of the RF amplitude
 produced by KTP  in the cavity (squares, right scale) for $I=7.1\,$MW/cm$^2$
and the 1st harmonic amplitude in the waveguide  (circles, left scale) for
$I=9.4\,$MW/cm$^2.$
    \label{fig:KTPconfrontocavitaguida}}
\end{figure} 
In \figref{fig:KTPconfrontocavitaguida} we compare the amplitude of the microwave produced by KTP irradiated in the
cavity with a laser intensity $I\approx 7.1 \,$MW/cm$^2$ (open squares) and
that of the 1st harmonic in the waveguide for $I\approx 9.4 \,$MW/cm$^2$.
The two sets of measurements are in good agreement. 

\begin{figure}[htbp]
    \centering 
\includegraphics[angle=180,width= \columnwidth]{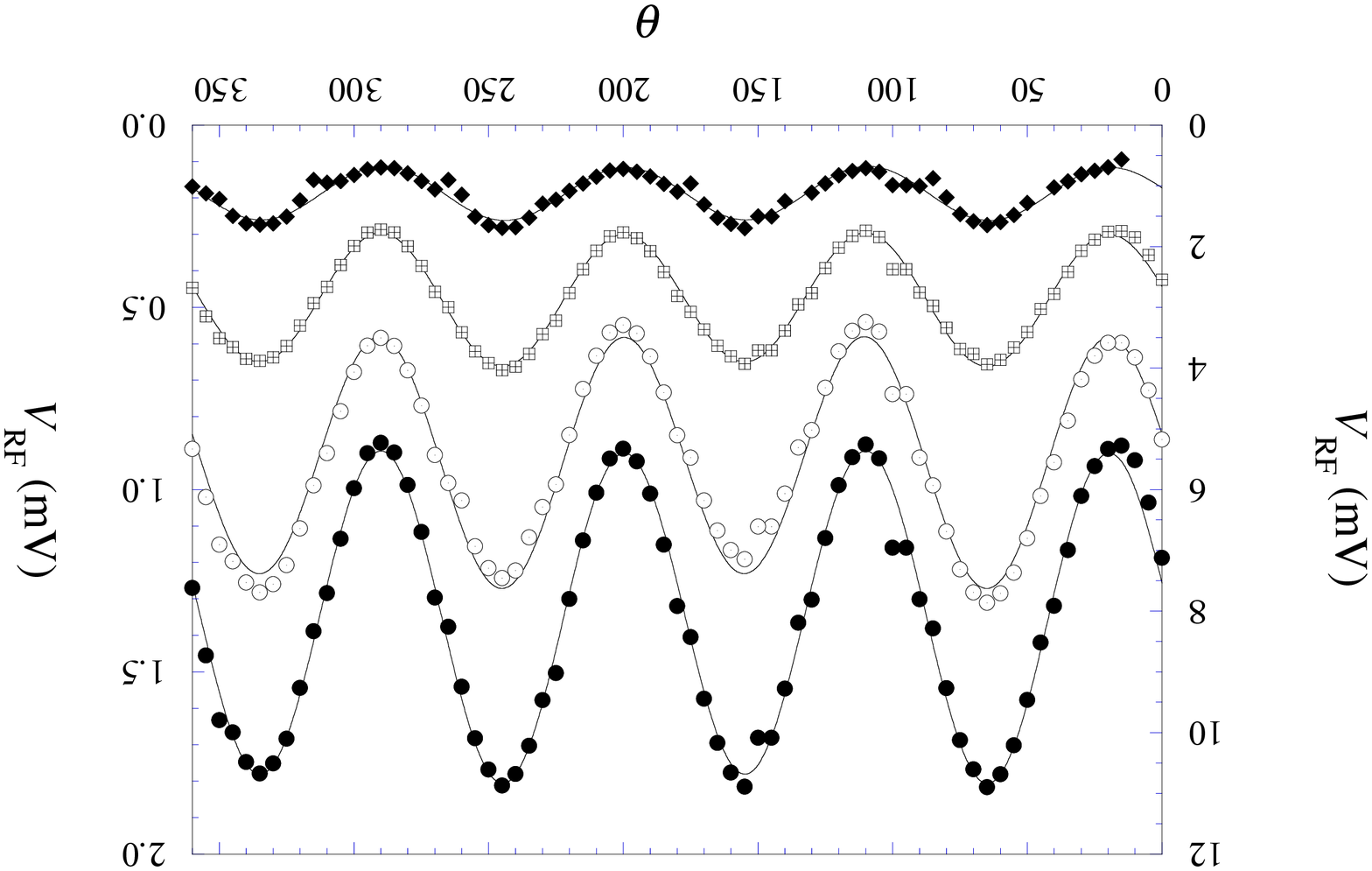}
    \caption{\small $V_{\mathrm{RF}}$ vs $\theta$ at constant $I=9.4\,$MW/cm$^{2}$
of the first four RF harmonics produced by KTP measured using the waveguide.
Left scale: 1st- (closed circle), 2nd- (open circles), and 3rd harmonic (squares).
Right scale: 4th harmonic (diamonds). The lines are only eyeguides.
    \label{fig:KTPguida4armoniche}}
\end{figure}

The dependence of the RF\ signal on $\theta$  is due to the function $h_i(4\theta,d_{ijk})$
and, according to~\eref{eq:d2p}, all RF\ harmonics should share the same
behaviour.
In \figref{fig:KTPguida4armoniche} we plot the angular dependence of the
first four RF harmonics of KTP measured in the waveguide at constant laser
intensity $I=9.4\, $MW/cm$^{2}$ and constant repetition rate $f_{0}=4.684\,$GHz.
As expected, the angular behaviour is the same for all of them.   Moreover,
it is easily verified that the four curves in \figref{fig:KTPguida4armoniche}
  coincide if each is multiplied by a suitable constant scaling factor.  The same
is true also for all other crystals. Once more, this observation remarks
the fact that all Fourier components of the RF spectrum share the same prefactor
$Ih_i(4\theta,d_{ijk})f_0^2.$

\begin{figure}[htbp]
    \centering 
\includegraphics[angle=180,width= \columnwidth]{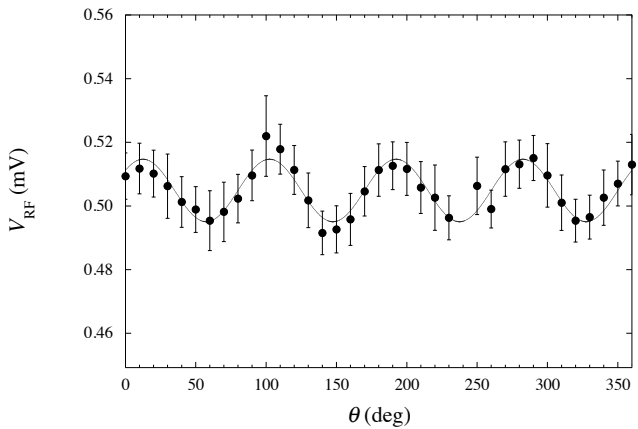}
    \caption{\small $V_\mathrm{RF}$ vs $\theta$ for LBO  for $I\approx 96.0\,$MW/cm$^2.
$ The solid line is the model prediction.
    \label{fig:LBO_VrfvsTheta}}
\end{figure}

\begin{figure}[htbp]
    \centering 
\includegraphics[angle=180,width= \columnwidth]{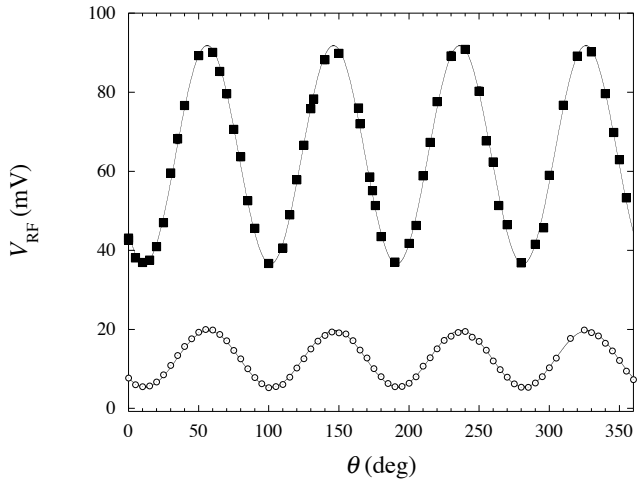}
    \caption{\small $V_\mathrm{RF}$ vs $\theta$ for KTP  for $I\approx 107.5\,$MW/cm$^2$
(squares) and for LiNbO$_3$ (circles) for $I\approx 45.0,$MW/cm$^2.$ The
solid lines are the model predictions. The dot size is comparable with the
error bars.
    \label{fig:LiNbO_KTP_VRFvsTheta}}
\end{figure}

\begin{figure}[htbp]
    \centering 
 \includegraphics[angle=180,width= \columnwidth]{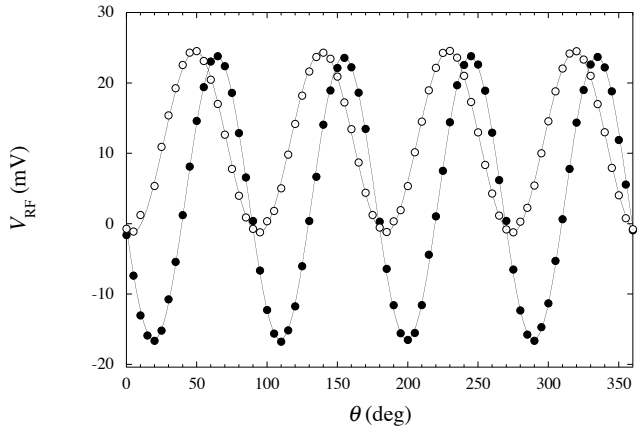}
    \caption{\small $V_\mathrm{RF}$ vs $\theta$ in ZnSe for $I\approx 8.1\,$MW/cm$^2$
for two different orientation of the crystal in the cavity. $\theta_c=205^\circ$
(closed dots) and $\theta_c = 115^\circ$ (open dots). The lines are the model
predictions. The
dot size is comparable with the error bars.
    \label{fig:ZnSe_Vrfvsth_thC205e115}}
\end{figure}

In \figref{fig:LBO_VrfvsTheta} through \figref{fig:ZnSe_Vrfvsth_thC205e115},
we show the angular dependence of the fundamental RF\ harmonic amplitude
measured in the cavity.  For all crystals,  $V_\mathrm{RF}$  shows  4-fold
periodicity as a function of   $\theta$ and, except ZnSe, the maximum RF\
production occurs at  the same angle of maximum SHG.  For ZnSe, the RF maximum
roughly occurs at the angle at which SH is minimum and vice versa. This behaviour
is related to the nonlinear tensor structure. That is why we carried out
RF measurements in ZnSe for two crystal orientations $\theta_c$ that differ
by $90^\circ$~(\figref{fig:ZnSe_Vrfvsth_thC205e115}). 

\noindent At constant $I$ and $f_0$,   the contribution of the second-order nonlinear
polarization to the amplitude of the RF signal measured in the cavity  is
given by the weighted sum of the contributions of each of its cartesian components
\begin{equation}
V_\mathrm{RF} =A\sum\limits_{i=1}^3 g_i h_i\left( 4\theta, d_{ijk}\right)
\label{eq:vrf}
\end{equation}                    
in which\ $A$ is a constant and $g_i$ are the director cosines. As the relative
orientation of the nonlinear polarization field relative to the cavity mode
polarization is unknown, the director cosines are to be treated as adjustable
parameters. 
 However, there is less freedom than it appears because
there is the strong constraint imposed by the crystal structure via the form
of the susceptibility tensor $d_{ijk}$  that determines the analytic form
of $h_i $ given by~\eref{eq:hi}, in which the components of the laser field
given by \eref{eq:epol} are expressed in the frame of reference of the crystallographic
axes. 

The  LiNbO$_3$ crystal is cut along a $(001)$ face so that the crystallographic
and the geometric axes coincide. We remind that the geometrical axes are
aligned to the laser polarization direction before  the half-wave plate.
LBO and ZnSe are cut along the $(01\bar 1)$ face so that the polarization
vector $\mathbf{\hat e}^\prime$ in the crystal frame of reference is obtained
by applying a rotation of amplitude $\alpha=135^\circ$ around $\hat x$
  \begin{equation}
\mathbf{\hat e}^\prime =\left(\begin{array}{c}\cos{2\theta}  \\ \cos{\alpha}\sin{2\theta}
  \\ -\sin{\alpha}\sin{2\theta}\end{array}\right)\label{eq:e1}
\end{equation}
The KTP crystal used in this experiment is cut at an angle $\beta\approx
165^\circ$ {  because it was previously used in a different experiment to obtain phase matching at a wavelength different from the present one~\cite{agnesi2005}}. In this case, the polarization
vector $\mathbf{\hat e}^\prime$ in the crystal frame of reference is obtained
as
 \begin{equation}
\mathbf{\hat e}^\prime =\left(\begin{array}{c}\cos{\beta}\cos{2\theta}  \\
\sin{2\theta}
  \\ -\sin{\beta}\cos{2\theta}\end{array}\right)\label{eq:e1KTP}
\end{equation}The solid lines in the figures are a fit to the data with a
function
of 4-fold
periodicity whose explicit analytic is specific to each crystal. For LBO, we have
\begin{eqnarray}
V_\mathrm{RF}&=&V_0  \bigg\{ -\frac{\sqrt{2}}{2}d_{15}g_x\sin{4\theta}+d_{24}g_y
\sin^{2}{2\theta} \nonumber \\
& &+ g_z \left[d_{31}\cos^{2}{2\theta} +\frac{1}{2} \left(d_{32}+d_{33}\right)\sin^2{2\theta}
 \right] \bigg\}  \label{eq:VrfLBO}\end{eqnarray}
The solid line in \figref{fig:LBO_VrfvsTheta} is obtained with fitting parameters:
$V_0 = 0.974\,$mV, $g_x=0.0122,$ $g_y = 0.835,$ and $g_z=0.550.$

For LiNbO$_3$ the fitting function is given by
\begin{eqnarray}V_\mathrm{RF}&=&V_0 \left[ g_x \sin{4\theta} +g_y \left(
d_{21} \cos^2{2\theta}
+ d_{22} \sin^2{2\theta}\right. \right)\nonumber \\
&+& \left. g_z \left(d_{31} \cos^2{2\theta}+d_{32}\sin^2{2\theta}\right)\right]\label{eq:VrfLiNbO}\end{eqnarray}
with values  $V_0= 4.47\,$mV, $g_x=0.561,$  $g_y=0.511,$ and $g_z=-0.652$
and is shown as the solid line in~\figref{fig:LiNbO_KTP_VRFvsTheta}. 

In the same figure, we plot the result of the analysis of the KTP crystal
whose fitting function is 
\begin{eqnarray}V_\mathrm{RF}&=&V_0 \bigg\{-2ab g_x d_{15}\cos^2{2\theta}
-bg_y d_{24}\sin{4\theta}+ g_z\times
\nonumber \\
&\times& \left(
a^2 d_{31}\cos^2{2\theta}+ d_{32}\sin^2{2\theta}+\right.\left. b d_{33} 
\cos^2{2\theta}
\right)\bigg\}
\label{eq:KTPfit}\end{eqnarray} 
in which $a=\cos\beta$ and $b=\sin\beta.  $ The fitting parameters are $V_0=37.9\,$mV,
$g_x=-0.502,$ $g_y=0.580,$ and $g_z=0.641.$

Finally, in \figref{fig:ZnSe_Vrfvsth_thC205e115} we show the results for
ZnSe for two by $90^\circ$ differing orientations of the crystal in the cavity.
The ZnSe crystal is also cut along the $(01\bar 1)$ face  and the three non
vanishing tensor elements are equal to each other so that the RF\ amplitude
has to be fitted to the function
\begin{equation}
V=V_0d \left[g_x \left(1+\cos{4\theta}\right)- \sqrt{2}\left(g_y + g_z\right)\sin{4\theta}\right]
\label{eq:fitZnSe}\end{equation}
in which $d$ is the common value of the non
vanishing tensor element. The solid lines in  \figref{fig:ZnSe_Vrfvsth_thC205e115}
are obtained with $V_0=33.5\,$mV, $g_x=0.360,$ $g_y=g_z= 0.660$ for the closed
symbols and $V_0=39.8\,$mV, $g_x=0.965,$ $g_y=0.031,$ and $g_z=0.262$ for
the open symbols. The different values of the director cosines are caused
by the rotation of the crystal within the cavity.\ The two determinations
of $V_0$ are compatible within the experimental accuracy.

\subsection{Dependence of SH generation on the laser polarization}\label{sect:SHG}

A byproduct of the measurements in the microwave cavity is the possibility
to observe SHG. As mentioned above, $V_\mathrm{G}$ is a quadratic function of the laser
intensity $I$ (see \figref{fig:VgvsIlallXTALSbis}). We note that~\eref{eq:ish} predicts that the SH intensity angular dependence
must be described by  functions of $8\theta$ and $4\theta$, only. The actual
form of these functions is related to the structure of the specific crystals.

If the crystals were perfectly aligned with the detector, the radiation impinging
on it would only originate from the two polarization components orthogonal
to  the line of sight. In the real experiment, the alignment is not perfect
so that the detector response at constant $I$ is a weighted sum of the contributions
due to all cartesian components of the nonlinear polarization.

\begin{figure}[htbp] 
    \centering 
\includegraphics[angle=90,width=0.6 \columnwidth]{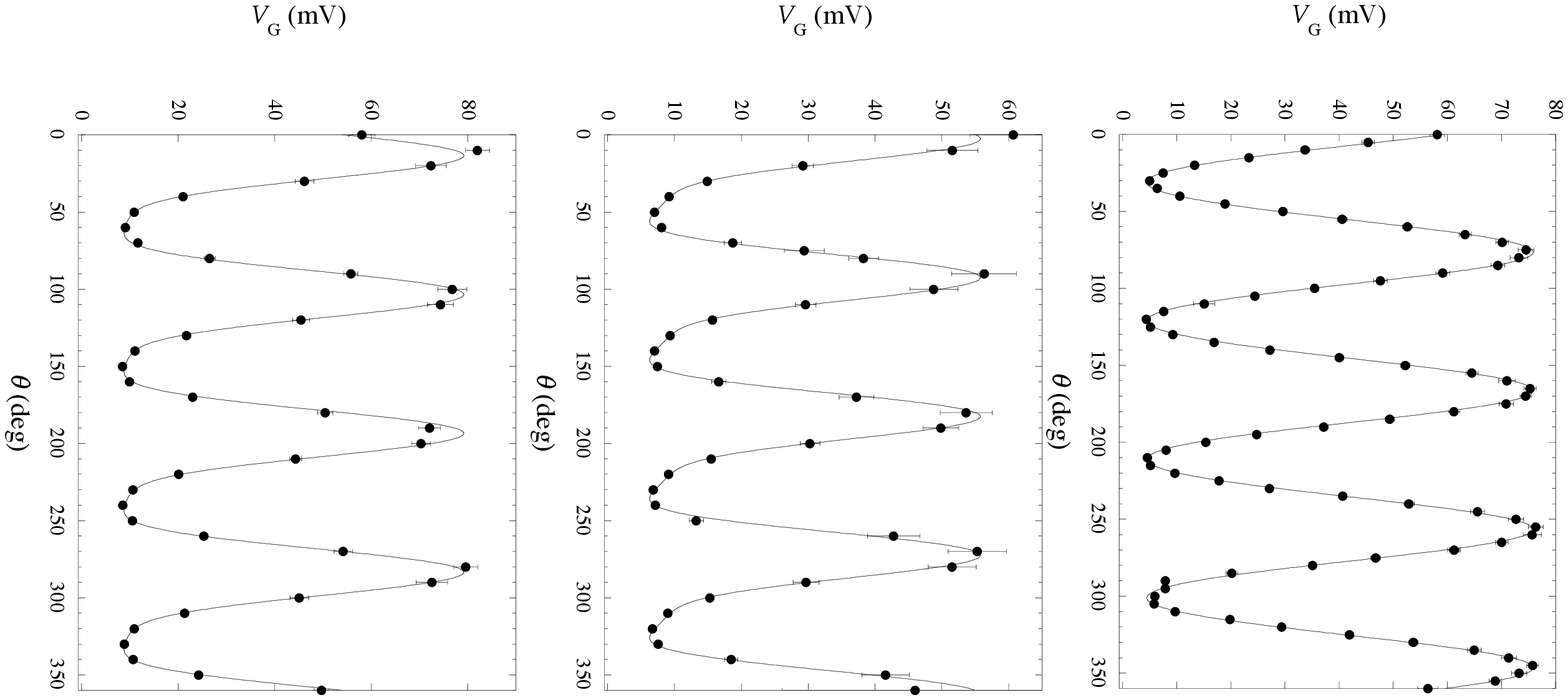}
    \caption{\small $V_\mathrm{G}$ vs $\theta$ for ZnSe (top) for $I=18.0\,$MW/cm$^2$,
LBO\ (middle) for $I=38.4\,$MW/cm$^2$, and LiNbO$_3$ (bottom) for $ I=11.5\,$MW/cm$^2$.
The solid lines are the predictions of the model. \label{fig:VGvsTheta_LBO_LiNbO_ZnSeL}}
\end{figure}

\begin{figure}[htbp] 
    \centering 
\includegraphics[angle=180,width= \columnwidth]{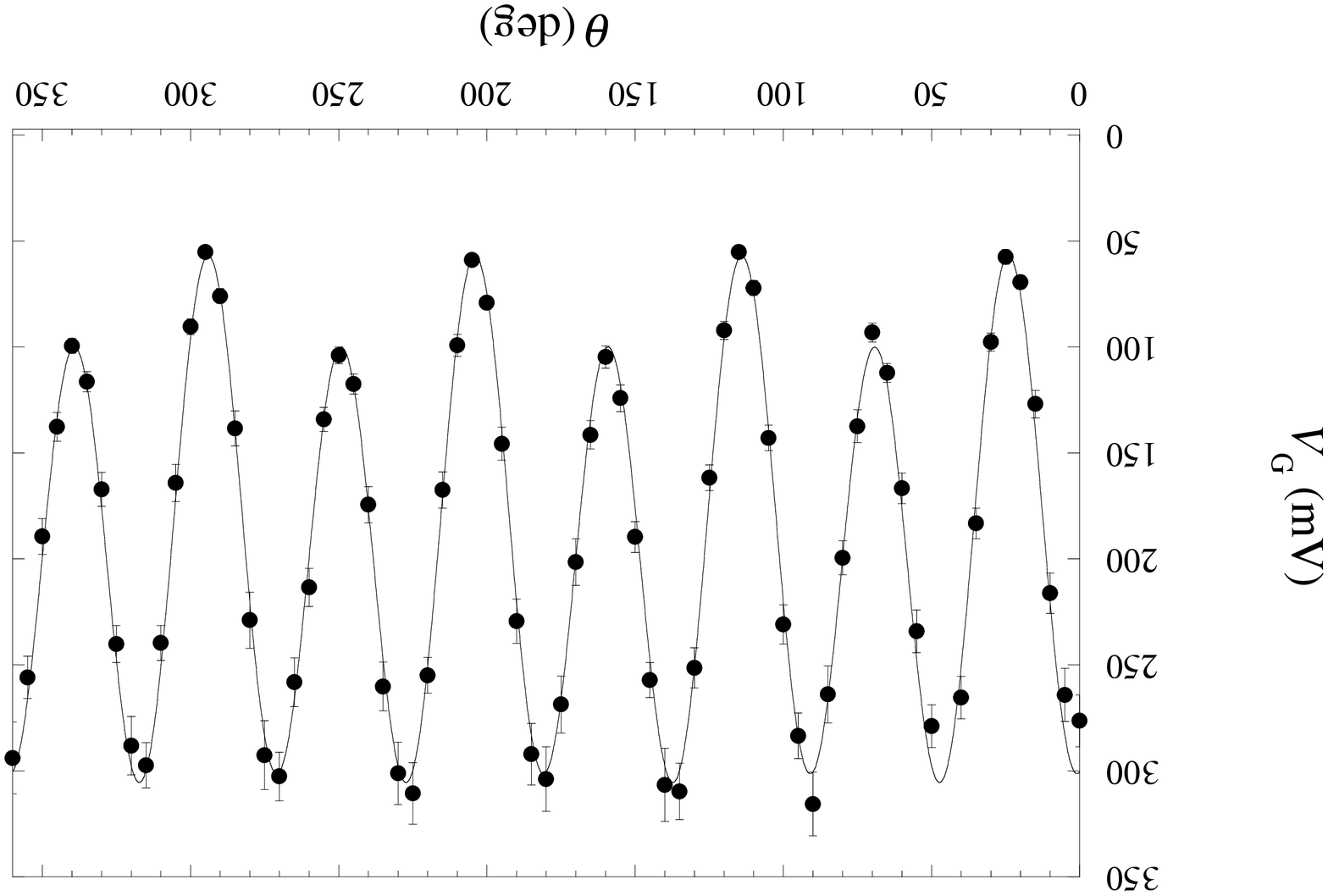}
    \caption{\small $V_\mathrm{G}$ vs $\theta$ for KTP for $I=119.7\,$MW/cm$^2$\cite{Borghesani2013}.
The solid lines are the predictions of the model.\label{fig:KTP_VGvsTheta}}
\end{figure}

 In~\figref{fig:VGvsTheta_LBO_LiNbO_ZnSeL} we show the SH intensity as a
function of $\theta$ for  ZnSe (top), LBO\ (middle) and LiNbO$_3$ (bottom)\  measured at relatively
low $I.$ In~\figref{fig:KTP_VGvsTheta} we report the SH intensity measured
in KTP~\cite{Borghesani2013} for the sake of comparison. The 8-fold periodicity
is very evident in KTP whereas it is hardly observable in ZnSe for which
the 4-fold periodicity is the dominant contribution. Also in LBO and LiNbO$_3$
the dominant contribution has a 4-fold periodicity and the 8-fold one manifests
itself only as a distortion of a nearly pure sinusoidal shape of the curves.

The  behaviour of the different crystals is once more related to the structure
of the second-order nonlinear susceptibility tensor. The behavior of KTP
has already been interpreted in our previous paper~\cite{Borghesani2013}
and its analysis will not be repeated here. 
According to ~\eref{eq:ish}, similarly to the RF\ analysis, the SH\ intensity
can be written as the weighted sum
\begin{equation}
V_\mathrm{G} (\theta) =\sum\limits_{i=1}^3 s_i v_i (8\theta,4\theta,d_{ijk})
\label{eq:Vgst}
\end{equation}
in which the analytic form of the functions $v_i$ depend on the crystal structure
and the weights $s_i$ are adjustable coefficients.

In ZnSe the SH\ intensity is a nearly pure sinusoidal function of 4-fold
periodicity. 
 Its fitting function, as the crystal  is cut along the $(01\bar 1)$ face
and as the three non vanishing elements of the nonlinear
tensor are equal $d_{14}=d_{25}=d_{36}\equiv d,$   takes
on the simple form
\begin{eqnarray}
V_\mathrm{G}&=& V_\mathrm{G0}\left(\frac{d}{4}\right) \left\{ \left( s_x
+\frac{3}{2} s_y + s_z\right)
+2 s_y \cos 4\theta\right. 
\nonumber \\
&+&  \left.\frac{1}{2}\left[s_y - 2\left( s_x+s_z\right)\right]\cos{8\theta}
           \right\} 
 \label{eq:ZnSeVg}\end{eqnarray}
The solid line in~\figref{fig:VGvsTheta_LBO_LiNbO_ZnSeL} (top) is obtained
with the parameter values $V_\mathrm{G0}= 9.71\,$mV, $s_x=s_{z}=0.250$ and
 $s_y =
0.935.$

For LBO, taking into account that the crystal is cut along the $(01\bar 1)$
face, the SH intensity is fitted to the function
\begin{eqnarray}
V_\mathrm{G}&=&V_\mathrm{G0}\left\{ -\frac{1}{\sqrt 2}d_{15}s_x \sin 4\theta
+  s_yd_{24}\sin^2{2\theta} + \right.
\nonumber \\
&+&\left .s_z \left[d_{31}\cos^2{2\theta} +\frac{1}{2} \left(d_{32}-d_{33}
\right)\sin^2
{2\theta}\right]\right\}
\label{eq:VGLBO}\end{eqnarray} 
We note that the LBO structure is such that only terms of  4-fold periodicity
appear in the fitting function. The solid line in~\figref{fig:VGvsTheta_LBO_LiNbO_ZnSeL}
(middle) is obtained with fitting parameters: $V_\mathrm{G0}=-77.9\,$mV,
$s_x=0.609,$ $s_y=0.110,$ and $s_z=-0.785.$ 
In LiNbO$_3$, the maximum SH intensity is obtained for a crystal orientation
different from that at which the RF amplitude is maximum.
The SH\ measurements are thus carried out by rotating the crystal by an angle
$\gamma $ around the $\hat y$ axis to roughly satisfy the phase matching
condition for maximum SH intensity. $V_\mathrm{G}(\theta)$ takes the form

\begin{eqnarray}
V_\mathrm{G}&= & V_\mathrm{G0}
\left[ s_x  \left(  -d_{15}  \sin{2 \gamma }\cos^2{2\theta} +d_{16} \cos{\gamma}\sin{4\theta}
\right)^2
 \right.\nonumber \\
&+& s_y \left( d_{21} \cos^2{\gamma}\cos^2{2\theta} -d_{24} \sin{\gamma}\sin{4\theta}
\right.
\nonumber \\
&+&\left. 
d_{22}\sin^2{2\theta} \right)^2  + s_z \left( d_{31}\cos^2{\gamma}\cos^2{2\theta}
\right.\nonumber \\
&+& \left.\left. d_{32}\sin^2{2\theta} + d_{33}\sin^2{\gamma}\cos^2{2\theta}\right)^2\right]
\label{eq:VgZnSe}
\end{eqnarray}
If $\gamma $ is left as an adjustable parameter, the solid line in~\figref{fig:VGvsTheta_LBO_LiNbO_ZnSeL}
is obtained for $\gamma \approx 19^\circ , $ which is very close to the value
$\gamma_c\approx 14^\circ$ of the phase matching angle at $\lambda=1064\,$nm\cite{weis1985}.
The remaining best fit parameters are $V_\mathrm{G0}= 5.67\,$mV, $s_x=0.911,$
$s_y=0.412 ,$ and $s_z=0.024.$

Finally, the previously published~\cite{Borghesani2013} KTP data in~\figref{fig:KTP_VGvsTheta}
are described by the function
\begin{eqnarray}
 V_{\mathrm{G}}&= &   V_{\mathrm{G0}} \left\{ 
 s_{x} \left(b d_{15}\right)^{2} \sin^{2}{4\theta} + s_{y}\left(  2abd_{24}\right)^{2}\sin^{4}{2\theta}\right.
 \nonumber\\ 
 &+&\left.  s_{z} \left[ d_{31}^{2}\cos^{4}{2\theta} +\left(a^{2}d_{32}\right)^{2}\sin^{4}{2\theta}\right.
\right. \nonumber\\
&+& \left. \left. \left( d_{33}b^{2}\right)^{2} \sin^{4}{2\theta}
 \right]\right\}
  \label{eq:}\end{eqnarray}
and the solid line in the figure is obtained with parameters $V_{\mathrm{G0}}=374.2\,
$mV,  $s_{x}=0.162,$ $s_{y}=-0.984, $ and $s_{z}=0.081.$

\section{Conclusions}\label{sect:conc} 
In this paper, we have investigated the efficiency of several second-order
nonlinear crystals to generate long microwave pulses when irradiated with
an intense pulse train of an infrared mode-locked laser because of optical
rectification. 
The spectrum of the emitted microwave radiation is a comb of integer multiples
of the laser repetition rate with a bandwidth determined by the single pulse
length that can  reach several tens of GHz.  By designing a suitable receiver
the desired microwave harmonic can be picked up and exploited for any given
use. 

We have also shown that the features of the microwaves generation depend
on the crystal structure. In particular, the RF emission efficiency depends
on the orientation of the crystal axes with respect to the laser polarization.
We have shown that the observed angular dependence of the RF amplitude agrees well with
 the theoretical prediction provided that the crystal orientation with respect to the receiver can be determined. 
 We have obtained a cross
check of this achievement by measuring the SH emitted by the crystals that
can be explained by the same model used for the microwaves. 

This technique can in principle be exploited to measure the elements of the
second-order nonlinear optical susceptibility tensor for unknown crystals
provided that great attention is paid to carefully control the geometry of
the experimental setup.
We finally note that an inline, nearly lossless device can be designed by
adopting this technique to monitor the quality of a high-repetition rate,
mode-locked laser, its only frequency limitation being determined by the
receiver used~\cite{Braggio2014}.

\ack
The authors wish to thank G. Bettella for supplying us the LiNbO$_3$ sample
and gratefully acknowledge helpful discussions with G. Carugno and the technical
assistance of E. Berto.

\section*{References}

\providecommand{\newblock}{}

\end{document}